\documentclass[conference]{IEEEtran} 
\usepackage[left=0.5in, top=0.5in, right=0.5in,bottom=0.5in ]{geometry}
\usepackage{graphicx,subfigure}
\usepackage{epsfig}
\usepackage{graphics}
\usepackage{cite}
\usepackage{subfigure}
\usepackage{balance}

\newtheorem{claim}{Claim}

\begin{document}
\bibliographystyle{IEEEtran} 
\title{Threshold Policy for Route Discovery Initiation\\in Mobile Ad hoc Networks}
\author{\IEEEauthorblockN{Tapas Kumar Patra and Joy Kuri}\\
\IEEEauthorblockA{Centre for Electronics Design and Technology\\
Indian Institute of Science, Bangalore\\
\{tkpatra,~kuri\}@cedt.iisc.ernet.in
}}
\maketitle

\begin{abstract}
Achieving optimal transmission throughput in data networks in a multi-hop wireless networks is fundamental but hard problem. The situation is aggravated when nodes are mobile. Further, multi-rate system make the analysis of throughput more complicated. In mobile scenario, link may break or be created as nodes are moving within communication range. `Route Discovery' which is to find the optimal route and transmission schedule is an important issue. Route discovery entails some cost; so one would not like to initiate discovery too often. On the other hand, not discovering reasonably often entails the risk of being stuck with a suboptimal route and/or schedule, which hurts end-to-end throughput. The implementation of the routing decision problem in one dimensional mobile ad hoc network as Markov decision process problem is already is discussed in the paper \cite{Patra07}. A heuristic based on threshold policy is discussed in the same paper without giving a way to find the threshold. In this paper, we suggested a rule for setting the threshold, given the parameters of the system. We also point out that our results remain valid in a slightly different mobility model; this model is a first step towards an `open' network in which existing relay nodes can leave and/or new relay nodes can join the network.\\
\end{abstract}

\section{Introduction }
\label{sec:intro}
Mobile multi hop ad hoc networks play a crucial role in setting up a network on the fly where deployment of network is not practical in times of utmost urgency due to both time and economical constraints. Industrial instrumentation, personal communication, inter-vehicular networking, law enforcement operation, battle field communications, disaster recovery situations, mobile Internet access are few examples to cite.

In Mobile Ad hoc network (MANET), communication between nodes situated beyond their radio range is also possible. For this type of communication, the nodes have to take help from other relay nodes which have overlapping radio communication. Here the communication is possible by knowing a path or route between the source and destination node. Transmission schedule should be known for the route. Finding the optimal route and transmission schedule shall be referred as `Route Discovery'. 

In static scenario route discovery is to be initiated only at beginning. In mobile scenario links may break or be created (nodes are moving within communication range).

We are motivated by the question: When to initiate route and schedule discovery in a MANET? A discovery entails some cost; so one would not like to initiate discovery too often. On the other hand, not discovering reasonably often entails the risk of being stuck with a suboptimal route and/or schedule, which hurts end-to-end throughput.

Our interest in this question stems from the need to assess how policies based on simple heuristics perform in comparison with policies that are optimal in some precisely defined sense. If it turns out that the simple heuristic is far from optimal, then the search for improved heuristics must continue. Else, it is reassuring to know that the heuristic performs nearly as well as it can.


In our earlier work~\cite{Patra07}, we had studied this problem in the framework of Markov Decision Theory. A simple one-dimensional network was considered, and a simple mobility model led to a Controlled Markov Chain, and our interest was in obtaining the best route and schedule discovery policy. The resulting problem was solved numerically, using the \emph{Value Iteration Algorithm} (VIA).

However, as pointed out in the earlier work, the VIA approach led to a huge computational burden. Computing the optimal policy required knowing the present `\textit{State}' (often impossible in practice), as well as significant computation. Therefore, a simple and suboptimal policy was considered: the \emph{Threshold Policy}. Whenever the end-to-end throughput dropped below a threshold, route and schedule discovery was initiated.

While the idea of a threshold policy is straightforward, the issue was the threshold value to use. In the earlier work, the best threshold was obtained by an exhaustive search within a finite set of possible thresholds: The one resulting in the best performance was found in this way. 
In this paper, we address this specific question: Can we arrive at a simple rule for setting the threshold, given the parameters of the system (number of relay nodes, number of positions, cost parameter, mobility parameters)? Even though the literature on MANETS is extensive, the issue of capturing the cost of route discovery in a formal framework does not seem to have received much attention. 

In this paper our contributions are:\\
i. Providing a rule that yields the threshold value for use in the threshold policy for deciding to do route and schedule discovery or not: Threshold value computed using the configuration information and ideal scheduling\\
ii. The study of the scheduling and end-to-end throughput characteristics which provides many incite of a linear ad hoc network.\\
iii. We also point out that our results remain valid in a slightly different mobility model; this model is a first step towards an `open' network in which existing relay nodes can leave and/or new relay nodes can join the network.\\
The boundary condition is relaxed and modeled as a wrap around condition for making a open end network. The mobility here is not necessary to be symmetrical. It is shown that the characteristic of the network does not change. 

Our results indicate that the performance of the proposed rule is no worse than 7\% of the best possible threshold threshold policy, and no worse than 15\% of the optimal, when the route discovery cost is low. 

In the following Section~\ref{sec:relatedWork}, the related work for this paper is discussed. In Section~\ref{sec:system-model}, the system model is described in detail. In Section~\ref{sec:recapitulation} discusses our previous work of finding long run average of  throughput which is studied in the framework of Markov Decision Theory. 
Section~\ref{sec:threshold-value}  discusses the derivation of threshold value for simple threshold-based heuristic and compare the performance with respect to the throughput-optimal policy. 
In Section~\ref{sec:OpenEndedBoundary}, we discussed the boundary conditions relaxed to make a open network. This network may cater the scenario which can be seen as a small area of concern in a large linear system. We conclude in Section~\ref{sec:conclusions}.
\section{Related Work}
\label{sec:relatedWork}
 Gupta and Kumar studied throughput of static wireless networks,~\cite{Gupta00}. They have considered protocol model and physical model for the studying of impact of interfering transmission on SNR. They observed that in a network comprising of $n$ identical nodes, each of which communicating with another nodes, the throughput per node under protocol model is of order $\Theta(\frac{1}{\sqrt{n~logn}})$ if placement of nodes is random. The throughput per node becomes $\Theta(\sqrt{n})$ if node placement and communication patterns is optimal. The later result is valid for physical model as explained intuitively by~\cite{JLi01}. While the overall on-hop throughput of the network grows as $\Theta(n)$, the average path length grows as $\Theta(\sqrt{n})$, which makes the throughput per node to vary as $\Theta(\frac{1}{\sqrt{n}})$. 

 Jain et al. used linear programming approach to characterize networks with interference, ~\cite{Jain03}. They used a conflict graph to model constraints on simultaneous transmissions. In the paper~\cite{Kodialam03}, both approximation algorithms that solve both the end-to-end flow routing problem and link scheduling problem near optimal are proposed. 
 In the paper~\cite{Arikan84}, it is shown that the problem of solving the optimal scheduling given the concurrency constraints to maximize network throughput, is NP-hard. 

In ~\cite{Tse02}, Grossglauser and Tse introduced mobility of nodes into the static model presented by~\cite{Gupta00}.

Many authors have discussed route discovery process ~\cite{Qiang08},~\cite{Fu05},~\cite{Banerjee10} but neither suggested when to initiate route discovery as their case is related to static case nor they have suggested how frequently to initiate the discovery process, in case of mobile network. Most of them suggested initiation of discovery only when a link, in the existing route, is disrupted i.e., in case of route break. In~\cite{Qiang08}, the authors proposed a modified AODV which uses the concept of reliable distance that change dynamically. Peng Fu et al.~\cite{Fu05} suggested distributed route discovery method that uses reinforcement learning. In~\cite{Banerjee10}, the authors have used fuzzy controller in every node. In their paper, the destination evaluates performance of all those routes and arranges it in order of preference, when route-request packet reaches its destination. 

In paper~\cite{Sakhaee07}, the authors discussed the route discovery initiation to reduce the frequency of flooding request by elongating the link duration of the selected paths.
In ~\cite{Reddy07}, the authors suggested extension in storing multiple paths as route rather than unipath as route .

In this paper  it is assumed that at first the techniques, to reduce the route discovery cost, are applied. And then, the discovery cost is represented as the fraction of the discrete time slot which will be explained in the Section~\ref{sec:system-model}. In this paper, we suggest a simple rule for setting the threshold value for the threshold policy, given the parameters of the system (number of relay nodes, number of positions, route discovery cost, mobility level), since the threshold policy is easy to implement. 

\section{System Model}
\label{sec:system-model}
\begin{figure}
	\centering
		\includegraphics[width=0.45\textwidth]{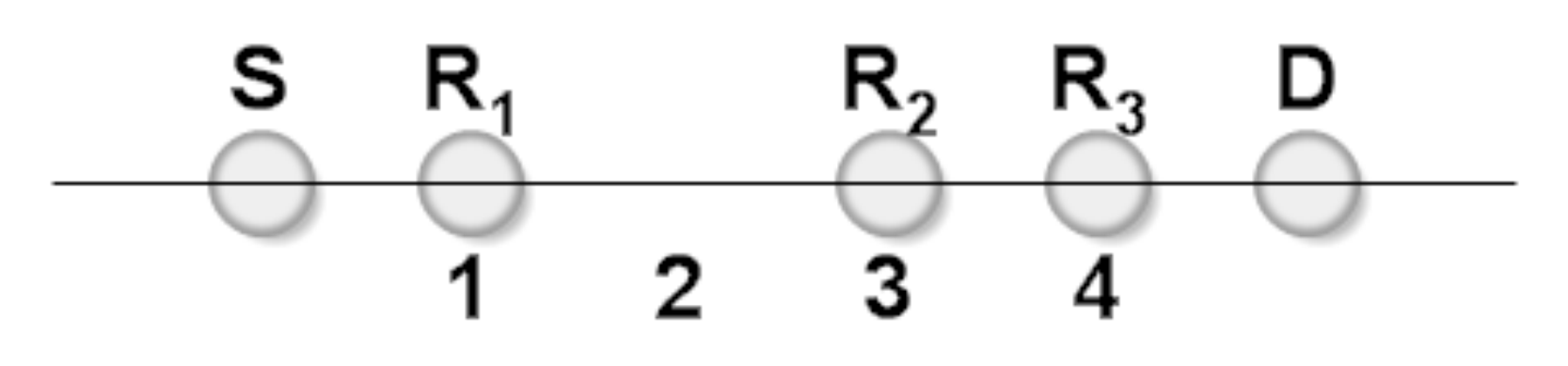}
	\caption{Linear mobile ad hoc network}
	\label{fig:model}
\end{figure}
We consider a network same as our earlier work in paper~\cite{Patra07} which is shown in fig~\ref{fig:model}. Here the network $G=(V,L)$ where $V$ is the set of vertices and $L$ is the set of links. While source and destination nodes are assumed to be fixed at the two ends of the linear grid, relay nodes are movable and can occupy any position in between the source and destination nodes. The number of possible positions which the relay nodes can occupy is $K$.
We will consider a bounded area i.e, number of the relay nodes is $N$ which is assumed to be same all the time. In the figure~\ref{fig:model}, the value of $K$ and $N$ are 4 and 3 respectively. Number of relay nodes $N$ can be more or less than the number of grid position $K$. But we will consider different ranges of node density as ratio of $N$ to $K$.

We will consider a discrete-time slotted system. Nodes can change grid position to left or right with a probability of $p_l$ and $p_r$ respectively only at the beginning of a time slot. A node may stay at the same position with probability $p_t=1-(p_l+p_r)$ at the beginning of the time slot but the node will not change the positions during whole period of time slot. However, if the node finds a boundary at the beginning of the time slot, it will wait at the boundary. We model the mobility of the network by specifying the duration of each time slot and the probability with which a node can move to the left or the right. Note that short (long) time slots correspond to a network with high (low) mobility. 

All nodes transmit and receive over a common channel. Transmission range is assumed to be $m$ units lengths of the linear grid. The link capacity or data rate will be 1 (normalized) if nodes are at neighboring positions, data rate will reduced to a smaller rate (1/2) if there is a vacant position in between and data rate will be 0 if there are at least two vacant places at two consecutive places. 

We assume interference range is more than the transmission range ($m$) but less than $m+1$ units of lengths i.e., if node transmits, it will interfere with any other nodes trying to transmit during the same time slot if the separation of nodes is less than interference range (here $m+1$). Again as any communication between two nodes requires exchange of packets both by the transmitter and by the receiver for setting up the link, both the nodes of a link should not be within the interference range of another communication link  at the same time. Among the links of communication between nodes if at least one of the nodes of each such links is in the interference range of others, only one link can be active at time. Hence the links whose both ends (the nodes) are away than both the nodes of other links, more than the interference range, can be active simultaneously. Hence for end-to-end communication through these links, the links are to be scheduled, i.e., when and what fraction of time they will be active satisfying the criteria discussed just now.

We model the \emph{cost} associated with route
discovery as follows. In every slot in which route discovery is initiated, we
assume that no data can be transmitted for a fraction $\phi$ of the
slot.

Suppose that route and schedule discovery takes no more than $\acute{t}$ units, where $\acute{t}$ is
less than a slot duration. Then the ratio of $\acute{t}$ to the slot duration is $\phi$.
Clearly, as $\phi$ moves closer to 1, the mobility level and the cost of route and 
schedule discovery increase. Correspondingly, as $\phi$ becomes smaller, the network 
is more and more static and the cost of route and schedule discovery can be amortized 
by sending more data over the slot. In the limit as $\phi$ goes to zero, we have a 
static network where route and schedule discovery is done at the beginning, and data 
can be transferred forever. This reduces to the model considered in, for example, \cite{Presti05}.

Just as $\phi$ is treated as a cost, the number of bits transferred over the slot duration 
behaves like a \emph{reward}. Suppose that an end-to-end transmission rate $R$ can be 
supported over the duration of the slot for the chosen route and transmission schedule.
Then, assuming that the slot duration is defined as the unit of time, the net reward 
over the slot is $(1 - \phi)R$ if route discovery is done.
When route discovery is not done, the net reward is simply $R$.
Clearly, the net reward corresponds to the number of data bits transmitted from the 
source to the destination during the slot.

A \emph{route} is defined as a sequence of grid-positions $(0, i_1, i_2, \ldots, i_l,\ldots, (K+1))$, where position 0 and $(K+1)$ indicate the positions of $S$ and $D$ respectively, and $i_1$, $i_2$, $\ldots$, $i_l$ indicate positions on the line, with $i_1 \leq i_2 \leq \ldots \leq i_l$, and $i_1 \leq m$, $(i_2 - i_1) \leq m$, $(i_3 - i_2) \leq m$, $\ldots$, $(i_l - i_{l-1}) \leq m$,  $((K+1) - i_l) \leq m$ in this one-dimensional network. 

Given a route, it is possible that there is no node at a particular position.
We still consider this as a valid route; however, the rate that can be supported on such  a route is clearly zero. Similarly, it is also possible that there are multiple nodes at a particular position. In this case, any of the nodes at that position can act as the relay node.
Because our performance criterion depends on the transmission \emph{rate} that corresponds to a route, the individual node identities do not matter.

\begin{figure*}[!ht]
  \centering 
 \subfigure[K=5, N=9]{\includegraphics[width=0.325\textwidth]{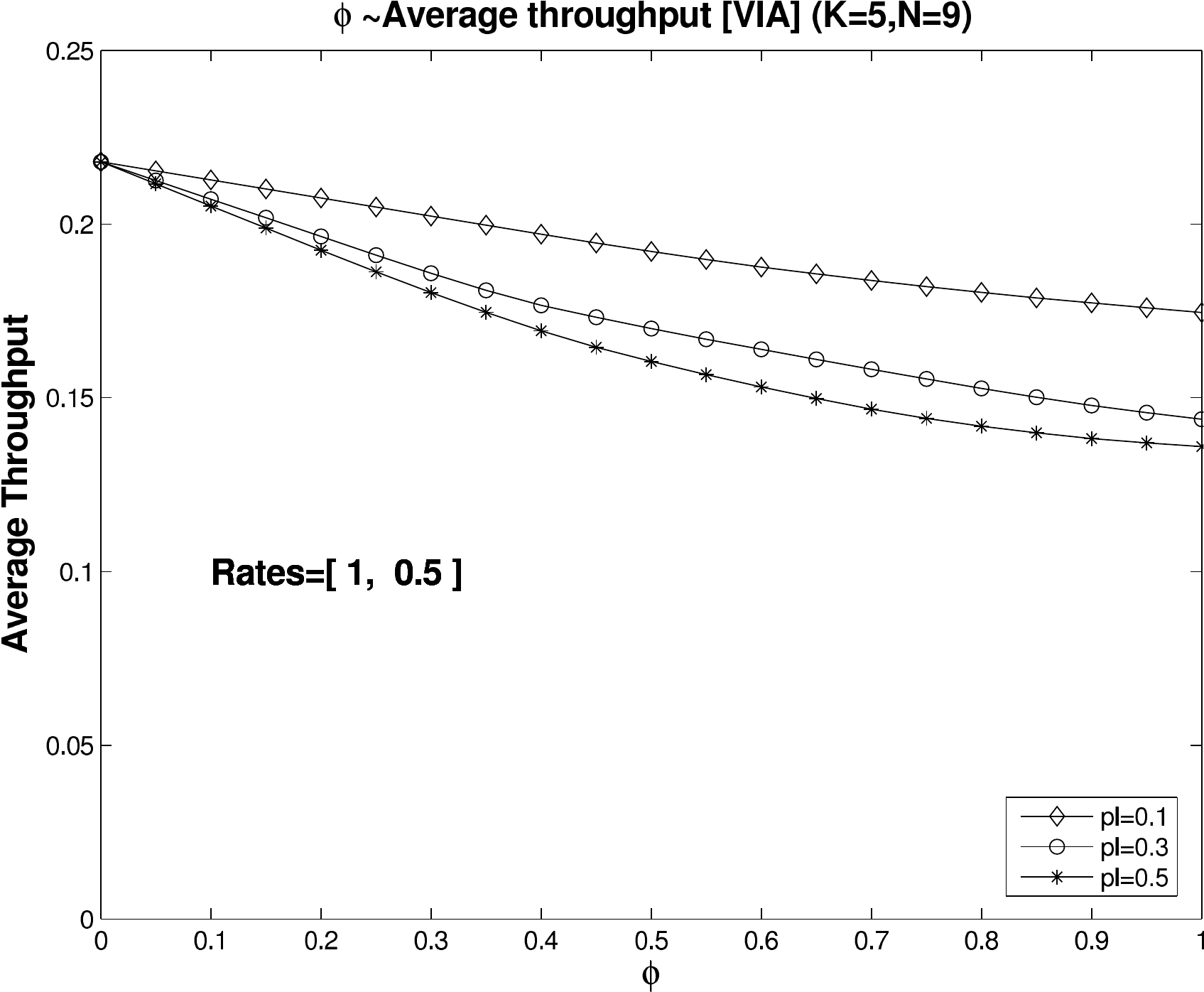}}
 \subfigure[K=6, N=3(y-scale enlarged)]{\includegraphics[width=0.325\textwidth] {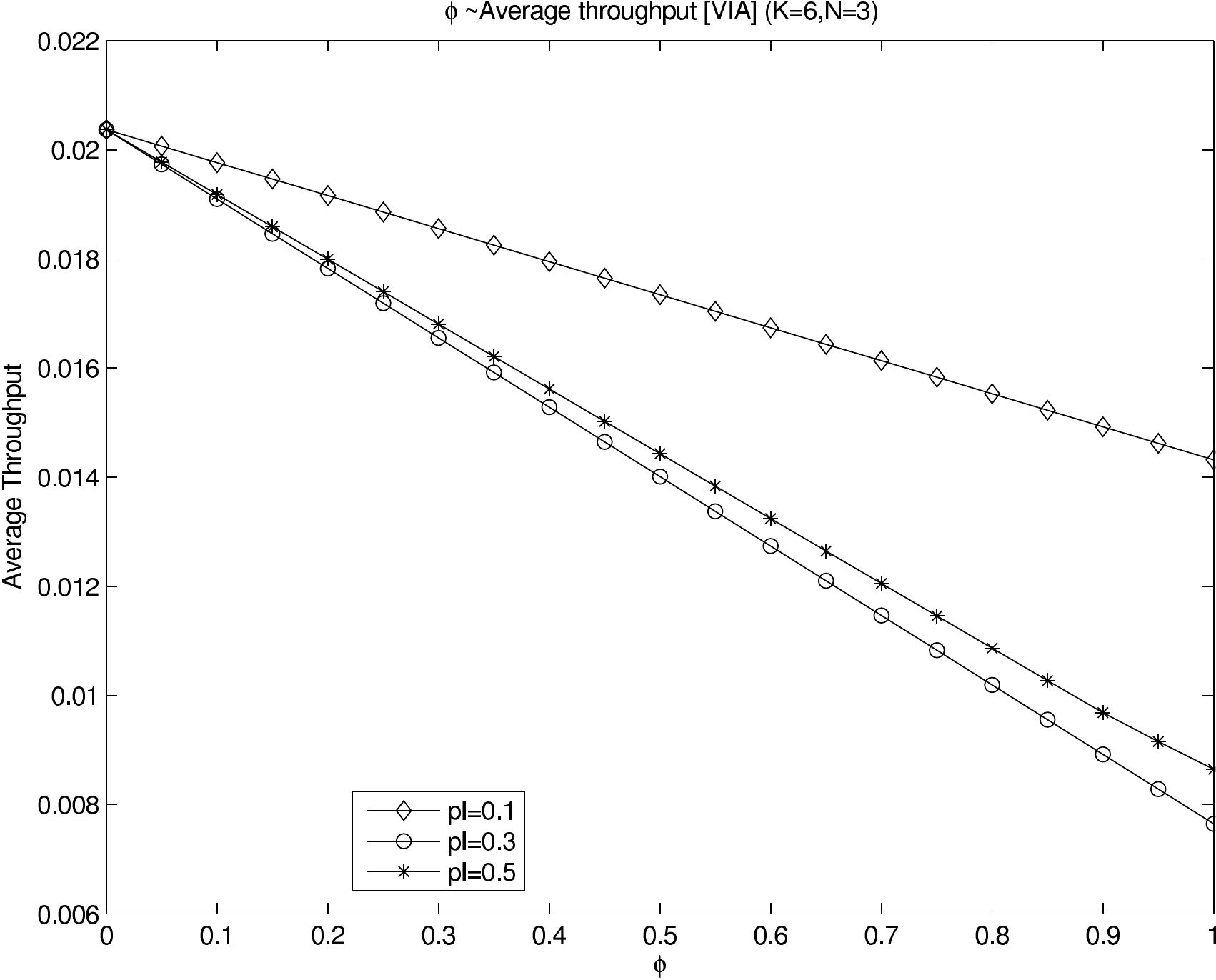} }
\subfigure[K=6, N=9]{\includegraphics[width=0.325\textwidth]{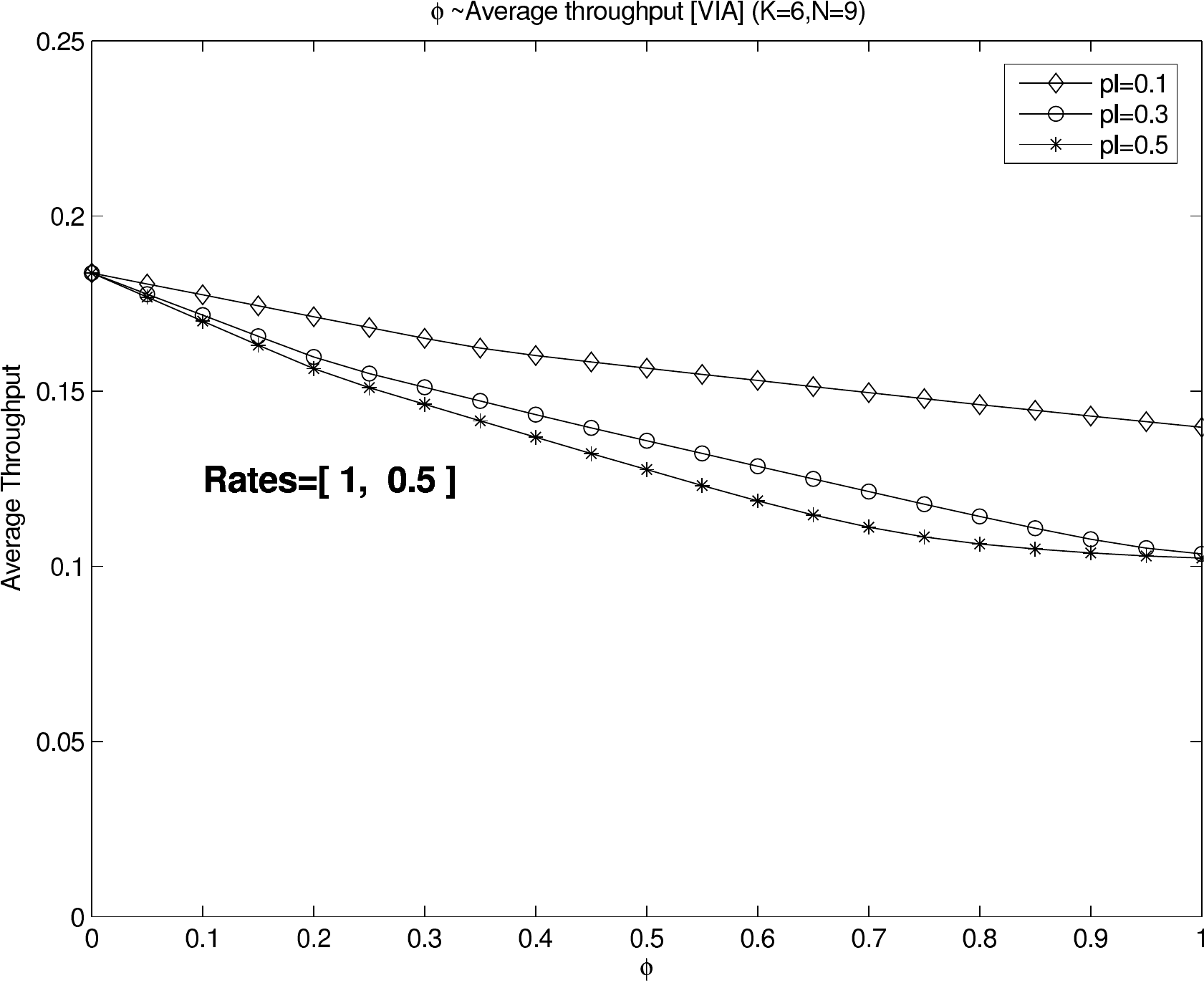}}
 \caption{Variation of optimal net throughput with $\phi$.}
  \label{fig:opt-thru-with-phi}
\end{figure*}
\section{Recapitulation of Earlier Work}
\label{sec:recapitulation}
The problem of route and schedule discovery was solved in the framework of Markov Decision Process (MDP)\cite{Sennott99} in our earlier work \cite{Patra07}. Five elements of MDP namely \textit{(a) the State Space}, \textit{(b) the Action Space}, \textit{(c) the Conditional Transition Probability given the current state and action}, \textit{(d) the One-step Expected Cost} and \textit{(e) the Total Cost Criterion over a finite or infinite time horizon}. The details of each elements can be found in the paper \cite{Patra07}. Some of the results are reproduced in Fig~\ref{fig:opt-thru-with-phi}. Two networks with $K=5,N=9$ and $K=6,N=3,9$ respectively are considered for the optimal net throughput with cost parameter($\phi$).

\begin{figure*}[!t]
  \centering
 \subfigure[$K=5, N=10, p_l=0.1$]{\includegraphics[width=0.325\textwidth]{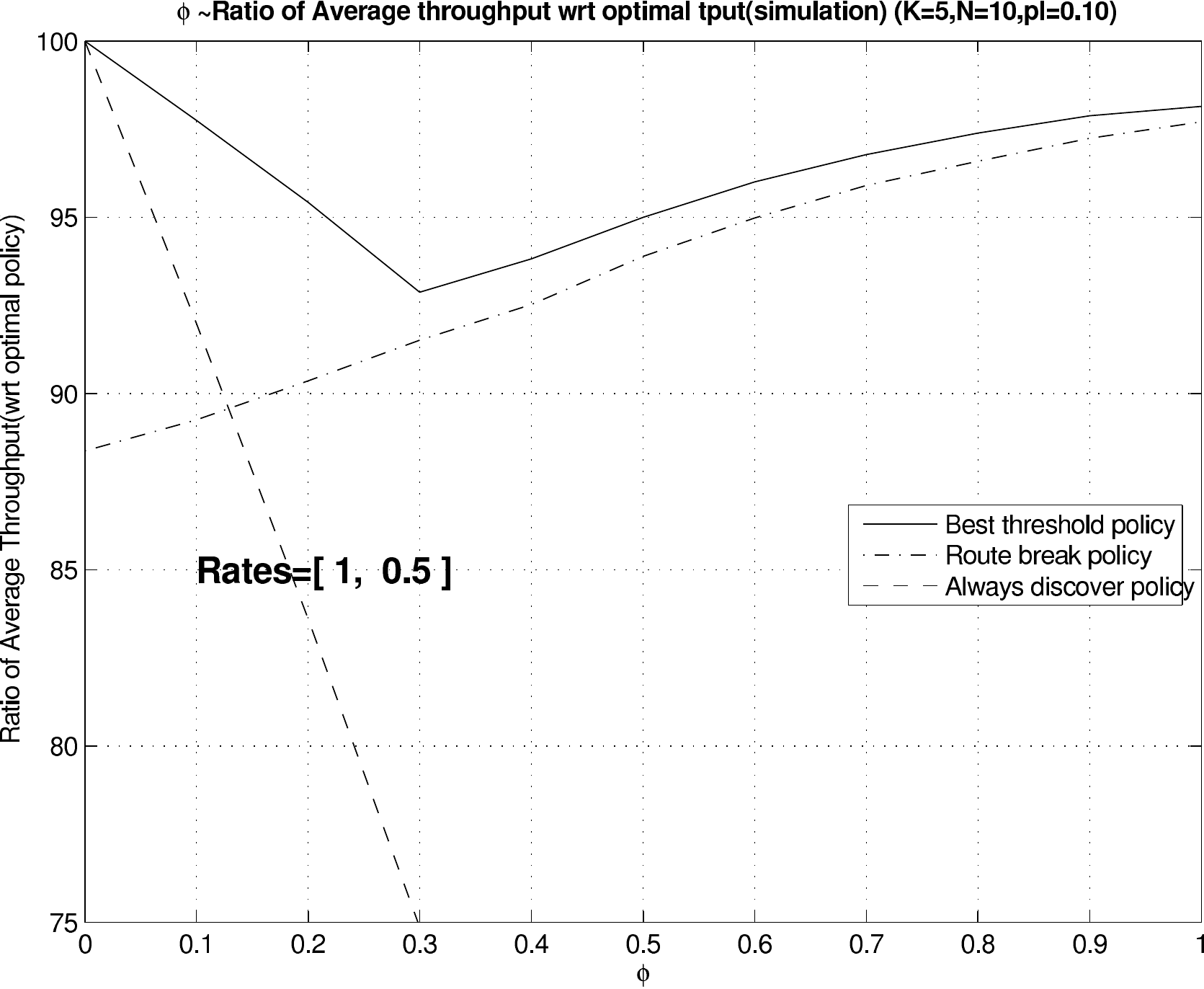} }
\subfigure[$K=5, N=10, p_l=0.3$]{\includegraphics[width=0.325\textwidth]{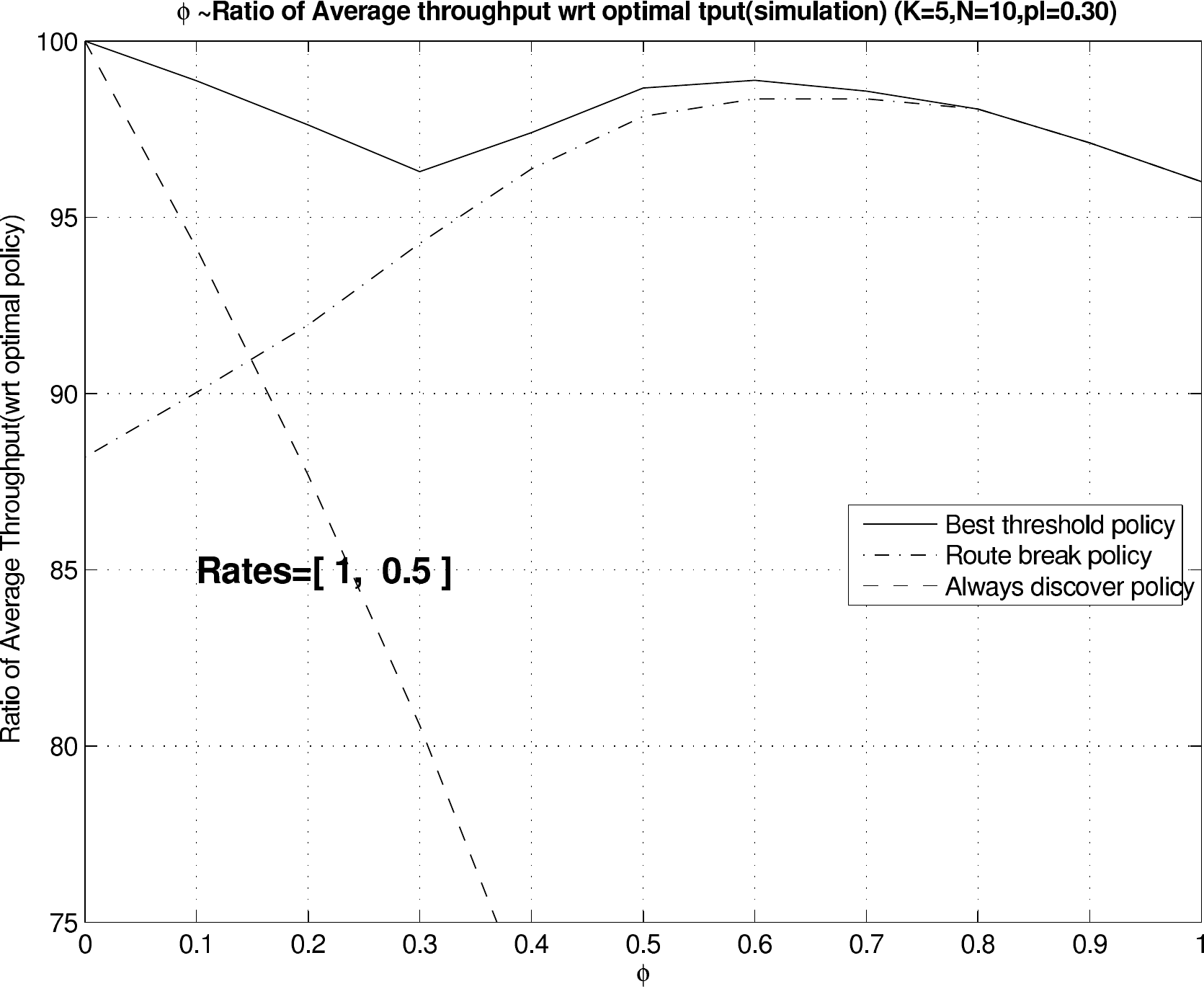}}
\subfigure[$K=5, N=10, p_l=0.5$]{\includegraphics[width=0.325\textwidth]{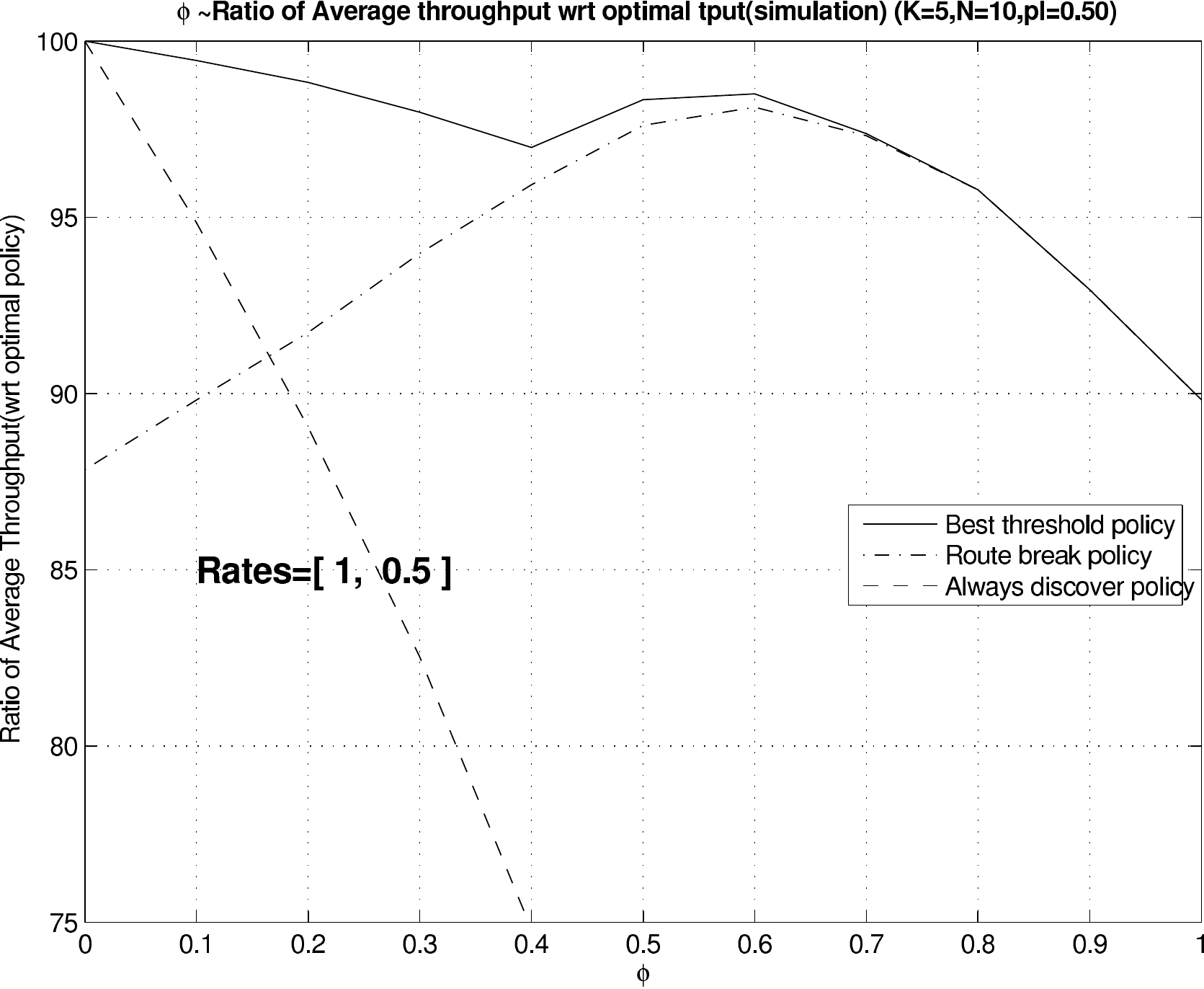}}
 \caption{Variation of average throughput of best threshold policy as percentage of average throughput of the optimal policy with $\phi$ and rates=[1 0.5].}
  \label{fig:efct-avg-tput-of-best-tput-pol-with-phiK5N10-05}
\end{figure*}

It is known that computing the optimal policy using VIA is a significant computational burden. Here we discuss a simplified policy which is used for obtaining high net end-to-end throughput. The motivation for this policy is as follows: If the observed throughput in a slot is small, then the current route is likely to be poor. The policy is: If the observed throughput is smaller than the threshold, then perform route and schedule discovery else continue with the currently known route and schedule. This is discussed in our earlier paper \cite{Patra07}. Some of the results are reproduced here in Fig~\ref{fig:efct-avg-tput-of-best-tput-pol-with-phiK5N10-05}. This figure indicates that, by proper choice of threshold value, there is advantage to implement the (best) threshold policy at the very low implementation cost. 

\section{Threshold Value }
\label{sec:threshold-value}
In this paper our objective is to find the threshold value which is close to the best threshold value that gives throughput as good as  the best threshold value. Also, we would like to compute this threshold value in a simple manner. Given a configuration, and ignoring any discovery cost, we can ask: What is the best possible throughput in this configuration? Let this end-to-end throughput named as `raw' throughput corresponding to a given configuration. Now allowing the configuration to vary over all possibilities, we can come up with an expected raw throughput. This is possible because we can find out the steady state probability of each configuration as given in Section~\ref{sec:ssprobability-configuration}.
\begin{eqnarray}&E(\mbox{raw throughput})=\ \ \ \ \ \ \ \ \ \ \ \ \ \ \ \ \ \ \ \ \ \ \ \ \ \ \ \ \ \ \ \ \ \ \ \ \ \ \ \ \ \ \ \ \ \ \nonumber\\
&\mbox{Steady state prob. of node distribution for the configuration\nonumber}\\
&\times\mbox{raw throughput for the best route.\ \ \ \ \ \ \ \ \ \ \ \ \ \ \ \ \ \ }
\end{eqnarray}
No discovery cost means $\phi$=0. Finally, we incorporate the role of discovery cost ($\phi$), by setting the threshold as follows. 
At higher $\phi$, the tendency to have less route discovery, with other conditions being same. Which implies that as $\phi$ increases, the threshold value will decrease. In other word, threshold value is decreasing function of $\phi$. \\We propose the following:
\begin{equation}
\label{eqn:threshold-value}
\mbox{Threshold Value}= (1-\phi^x)*E(\mbox{raw throughput}), x>0	
\end{equation}
\subsection{Steady State Probability of Specific Configuration (States based on Positions of users ) }
\label{sec:ssprobability-configuration}
It can be easily shown that the positions of single node when movement is random walk with boundary behavior `pause and restart (stuck-at-boundary)' in one dimension, is uniformly distributed at all movement positions. This is because of doubly stochastic nature of the state transition probability matrix.\\
The probability that the node is at any of the moving positions $=\frac{1}{K}$ and when $N$ such nodes are there, steady state probability of any ordered configuration of $N$ nodes =$(\frac{1}{K})^N.$\\
As in our case the first part of the state which is based only on movement positions counted as `how many nodes are at one movement positions' with $K$ such movement positions, we have to find out how many numbers of ordered pair of nodes those make one single state as discussed.\\
This problem reduces as follows:\\
 Let there be $[n_1,n_2,...,n_K]$ nodes at the $K$ grid positions respectively. Hence $\sum_{i=1}^{K}{n_i}=N$.\\
 It can be shown that for the $k^{th}$ position, the number of possible options is $A_k=^{(N-\sum_{i=1}^{k-1}{n_i})}C_{n_k}$.\\
Hence the total numbers of ordered pairs of nodes that make one single state is 
\begin{eqnarray}
 &=&\prod_{i=1}^{K}A_i=\prod_{i=1}^{K-1}A_i\\
&=&\frac{N!}{n_1!(N-n_1)!}\times \frac{(N-n_1)!}{n_2!(N-n_1-n_2)!}\times... \\
&=&\frac{N!}{n_1!...n_K!}
\end{eqnarray}
Now the steady state probability for the state $[n_1,n_2,...,n_K]$ is
\begin{eqnarray}
\label{eqn:ssProbability}
&=&\frac{N!}{n_1!...n_K!}\times (\frac{1}{K})^N.
\end{eqnarray}
\subsection{Best Throughput}
For any configuration, there will be at most a certain number of routes possible according to $K$ and $N$. For ex. with $K=4$ as in Fig~\ref{fig:model}: if route$(S-1,1-2,2-3,3-4,4-D)$ is represented as $(S,1,2,3,4,D)$ then the possible routes are $(S,1,2,3,4,D), (S,1,2,3,D), (S,1,2,4,D), (S,1,3,4,D), \\(S,1,3,D), (S,2,3,4,D), (S,2,3,D), (S,2,4,D)$ and (null route). For each route, the best throughput can be computed by using optimal scheduling as discussed in Section~\ref{sec:opt-sch-ind-set}. The null route is added here just to address the situation when the system has no route at the beginning.
\subsection{Optimal Scheduling}
\label{sec:opt-sch-ind-set}
This part of the our derivation is similar to the derivation of the problem when the network is static as in paper~\cite{Presti05}.
Any communication between two nodes causes contention with any other node within the interference range of both the nodes if both are active simultaneously. This problem is approached using `Conflict graph' whose vertices correspond to the links of the transmission graph ($G$) of the network. In this conflict graph an edge from a vertex to itself is not drawn. If the edge between two nodes exist then the corresponding links in transmission graph interfere with each other and hence can not be active simultaneously. 

Links belonging to an independent set in the conflict graph can be scheduled simultaneously. Using maximal independent set the optimal scheduling problem can be expressed as linear program. And solving the linear program, we can get link schedule i.e., the fraction of time the links which will be active. 
\subsection{Results}
\begin{figure}
	\centering
	\includegraphics[width=0.49\textwidth,height=0.25\textwidth]{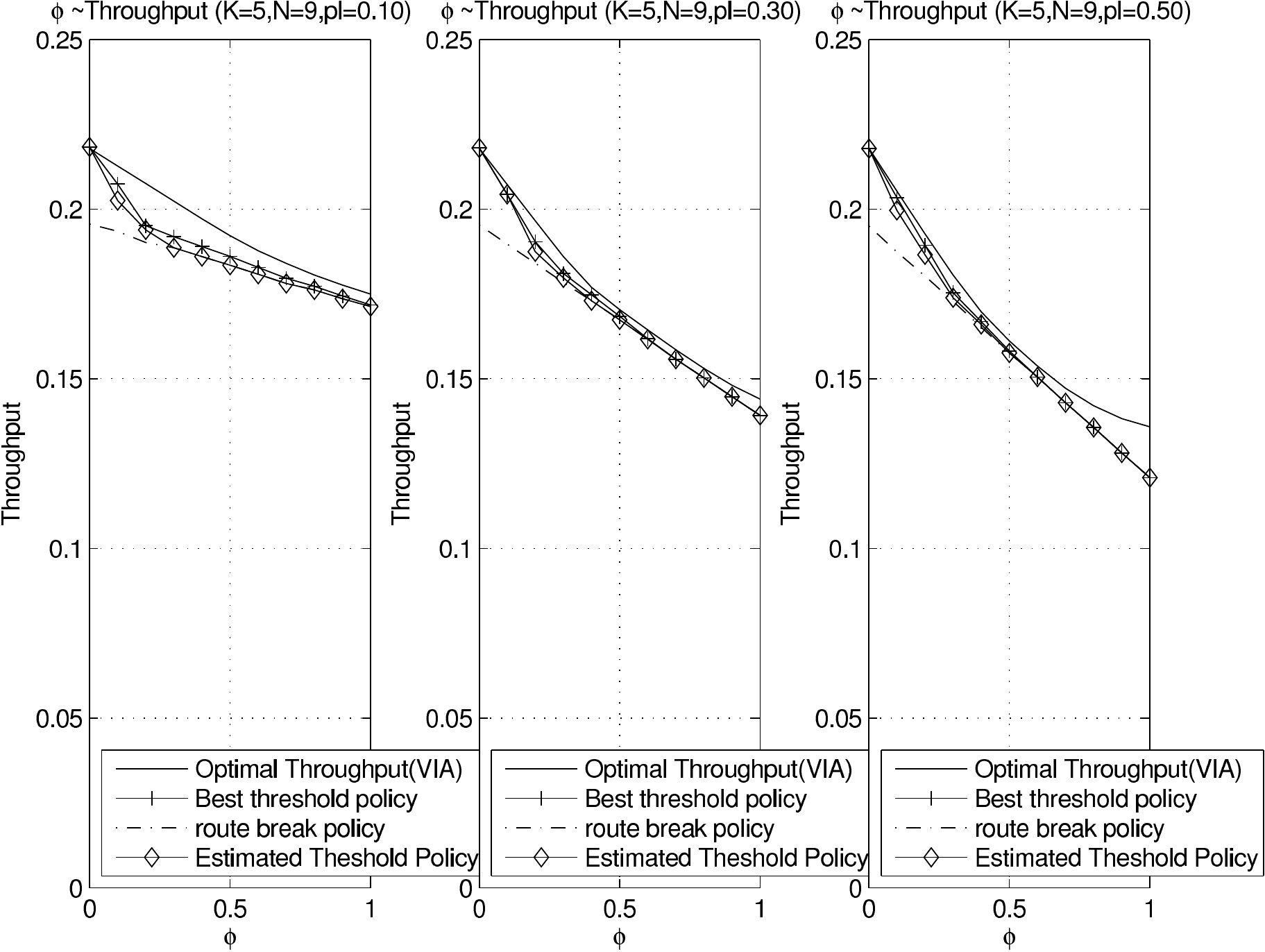}
	\caption{Throughput vs. $\phi$ when the expected throughput is as per the rule as in eqn ~\ref{eqn:threshold-value}}
	\label{fig:grEXTh5c1K5N9}
\end{figure}

\begin{figure}
	\centering	\includegraphics[width=0.49\textwidth,height=0.25\textwidth]{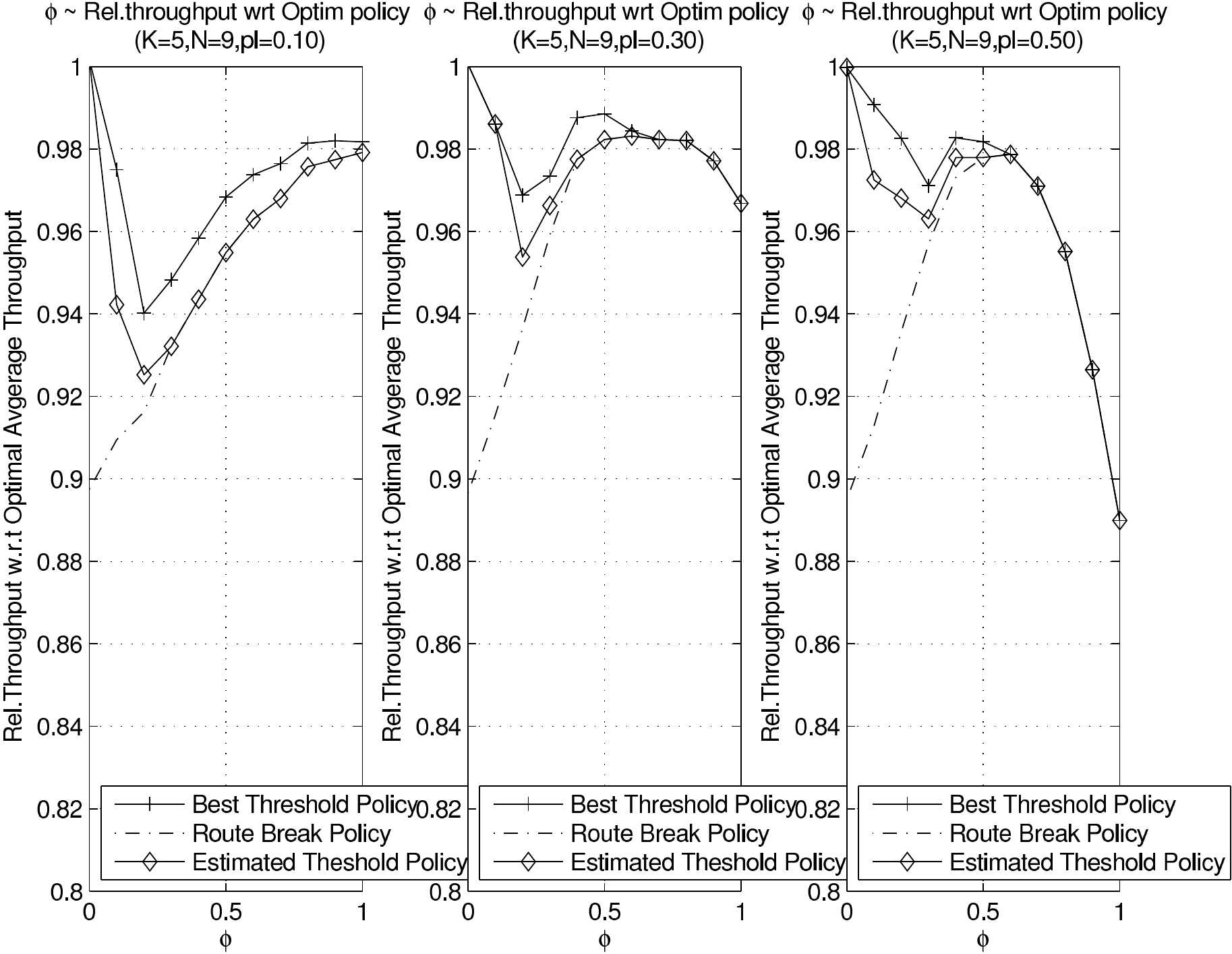}
	\caption{Relative throughput with respect to optimal policy vs $\phi$}
	\label{fig:grEXTh5jK5N9_wrtOptim}
\end{figure}

The expected raw throughput computed using the above method is taken as a threshold value for the threshold policy and simulations are done for different system parameters. From simulations it is observed that $x=2$ gives a good approximation for most of the cases. The related graphs are given in Fig~\ref{fig:grEXTh5c1K5N9},~\ref{fig:grEXTh5jK5N9_wrtOptim}. It can observed from the graphs that most of the configurations, when $\phi\le 0.5$, the performance will not be worse than 7\% of the best possible threshold value and will not be worse than 15\% of the average throughput as obtained by using the optimal policy. As $\phi> 0.5$, it is observed that threshold policy follows the route break policy. 
\section{Open Ended Boundary}
\label{sec:OpenEndedBoundary}
The boundary condition explained earlier is close ended system i.e., Stuck-at-boundary model is necessarily make the number of relay nodes in the area of concerned constant as nodes are neither allowed to leave or join the existing network. But the model is meaningful only when mobility is symmetrical i.e., $p_l=p_r$; otherwise, eventually, all the nodes move to the leftmost or rightmost position with probability 1. To allow the unsymmetrical mobility model($p_l\ne p_r$), another boundary model namely \textit{wrap-around} model is considered. To make the relay node constant, an assumption, though little bit artificial, is made: if a node move out of(into) the area at one end then another node is move into(out of) the area at the other end.
\subsection{Wrap Around Boundary Conditions (Open Ended Boundary). }
\begin{claim}
The fraction of time a single node is at any of the moving positions for 1-dimensional random walk with \emph{ wrap around} boundary conditions is uniformly distributed even when moving probabilities toward left or right are unequal.
\end{claim}	
\IEEEproof 
In this model, when ever a boundary is found, instead of jumping out of the bounded area, node will be transfered to the other end for that time slot.
\begin{math}
\mbox{The state transition probability matrix (P)}=\\
\bordermatrix{&1		&2	 &3		&4		&...& K\cr
          1 & p_t  & p_r  & 0   &...&...  & p_l  \cr 
          2 & p_l        & p_t  & p_r & 0&...&0  \cr
          3 & 0     & p_l     & p_t  & p_r  &...&0 \cr
          ... & ... & ...  & ...  & ...&... &...  \cr 
          K & p_r  & ... &...&0 & p_l & p_t \cr}
\end{math}\\
 As the state transition matrix is doubly stochastic matrix, even when mobility is non-uniform,  the steady state distribution ($\pi$)=$[1/K,...,1/K,...,1/K]$. So the earlier analysis applies.

\section{Conclusions}
\label{sec:conclusions}
Threshold policy is a practical method for being both a simplified, less computational intensive approach, and the approach which can be implemented by measuring the throughput instead of knowing the states. The policy that measures the throughput systematically along with randomly measuring it relieved from the requirement to know the mobility condition (time slot depends upon mobility) by measuring the change of throughput also. The expected throughput is analytically derived in this paper and it is observed that $(1-\phi^2)$ is a good approximation to the multiplying factor for most of the cases when $\phi$ is considered. The analytical method for finding the steady state probability of different configurations based on number of relay nodes at different nodes is obtained. It is observed from the simulations that for most of the configurations, by considering the stated rule, when $\phi\le 0.5$ the performance of the proposed rule is no worse than 7\% of that of the best threshold policy, and no worse than 15\% of the optimal. As $\phi> 0.5$, it is observed that threshold policy follows the route break policy.

The the boundary condition is modified to satisfy open network where nodes can leave/join the network. This is modeled as wrap-around boundary condition. Here assumption taken is: whenever a node leaves (joins) another node joins (leaves) simultaneously at the other end of the boundary so as to make the number of nodes in the network same. It is analyzed that the behavior in this case is also same as Struck-at-boundary condition. `By withdrawing the previous assumption, i.e., the number of relay nodes varies randomly, now the analysis can be modeled as birth-death process', is the future work we are continuing now.

\end{document}